\documentclass[12pt]{article}

\usepackage{amsmath}
\usepackage{amsfonts}

\begin{document}
\thispagestyle{empty}
\begin{centering}
\large{Final arXiv version of}
\\ \vspace{0.3in}
\Large{Comment on Zhang, D. Exact Solution for\\
Three-Dimensional Ising Model.\\
Symmetry 2021, 13, 1837}\\
\large{(published as Symmetry 2023, 15, 374
[arXiv:2110.11233])}
\\ \vspace{0.3in}
followed by a response to
\\ \vspace{0.3in}
\Large{Reply to Perk, J.H.H. Comment on ``Zhang, D.\\
Exact Solution for Three-Dimensional\\
 Ising Model. Symmetry 2021, 13, 1837''}\\
 \large{(published as Symmetry 2023, 15, 375
 [arXiv:2302.10139])}
\\ \vspace{0.3in}
\Large{Jacques H.H.\ Perk\\
Department of Physics\\
Oklahoma State University\\
Stillwater, OK 74078-3072, USA}\\
\end{centering}

\newpage\pagenumbering{arabic}
\title{Comment on ``Exact Solution for Three-\\
Dimensional Ising Model'' by Degang Zhang\\
{\Large followed by}\\
Response to the Reply by Degang Zhang}

\author{Jacques H.H.\ Perk\\
Department of Physics, Oklahoma State University\\
Stillwater, OK 74078-3072, USA}
\maketitle

In \cite{ZhangDG}, Zhang Degang claims to have solved the
free energy per site of the three-dimensional Ising model
with screw boundary conditions.
This claim evolved in earlier versions \cite{ZhangDG0}, in part due
to my referee reports on them, in which I stated that I had found
that the claimed result failed the well-known high-temperature series
test for the thermodynamic free energy. This failure was confirmed
in version 4 of \cite{ZhangDG0}, leading to the additional erroneous
claim of the dependence on boundary conditions of the free energy
per spin in the thermodynamic limit.

However, this free energy per site in the thermodynamic limit
is to be independent of boundary conditions \cite{Griffiths,Ruelle}.
This follows from the Peierls--Bogolyubov inequality, which implies
that the difference of total free energies is bounded
by the norm of the difference of their Hamiltonians,
\begin{equation}
\big|F[\mathcal{H}_2]-F[\mathcal{H}_1]\big|
\le ||\mathcal{H}_2-\mathcal{H}_1||,\quad
F[\mathcal{H}_i]\equiv-\beta^{-1}\log\mathrm{Tr}\,\mathrm{e}^{-\beta\mathcal{H}_i},
\;(i=1,2).
\end{equation}
As
$ln$ (or $2ln$ according to Figure 1 in \cite{ZhangDG})
bonds $J_1$ are moved from\linebreak periodic to screw boundary
conditions, the right-hand side of (1) is bounded by
$2ln|J_1|$ (or $4ln|J_1|$). Per site, we must divide by $lmn$
(or $2lmn$), so the free energies per site differ by at most
$2|J_1|/m$, which becomes zero in the\linebreak thermodynamic limit
$l,m,n\to\infty$. Hence, the results per atom for periodic
and screw boundary conditions, and also (44) and (45) in
\cite{ZhangDG}, should be equal.

On the middle of page 5 of \cite{ZhangDG}, we read
$\mathcal{A}_{p,s}\equiv A_{p,s}$, equivalent  because they
commute and have the same eigenvectors and eigenvalues.
However, equating this ``$\equiv$'' and ``='' is a serious error, as then
$\sigma^z\equiv-\sigma^z$ would imply $\sigma^z=-\sigma^z$,
for example.
More generally, we should expect the common eigenvalues of $\mathcal{A}_{p,s}$
and $A_{p,s}$ to be distributed differently over the common eigenvectors, so that
$\mathcal{A}_{p,s}\ne A_{p,s}$ instead of being equal.

This error is present in (16) in \cite{ZhangDG}, where we may replace
$\sum_{p=1}^m \mathcal{A}_{p,1}=\sum_{p=1}^m A_{p,1}$ by
$\mathcal{A}_{p,1}=A_{p,1}$. (As $m$ is arbitrary, the equality of the
two sums in (16) is equivalent to the equality of their summands.)
However, by (15) and the text below
it in \cite{ZhangDG},
\begin{eqnarray}
\mathcal{A}_{p,1}&=&\mathcal{L}_{1,2}^p=\sigma^z_p\sigma^z_{p+m}
\equiv A_{p,1}=L_{1,2}^p\nonumber\\
&=&L_{p,p+m}=
\sigma^z_p\sigma^x_{p+1}\cdots\sigma^x_{p+m-1}\sigma^z_{p+m},
\end{eqnarray}
after both  reconstructing missing definitions from the text below (13)
and using (3) in~\cite{ZhangDG}. Indeed, $\sigma^z_p\sigma^z_{p+m}$
and $\sigma^z_p\sigma^x_{p+1}\cdots\sigma^x_{p+m-1}\sigma^z_{p+m}$
commute and have the same eigenvectors and eigenvalues, but these $\pm1$
eigenvalues are distributed differently over the common eigenvectors. Therefore,
equating $\mathcal{A}_{p,1}=A_{p,1}$, as is used in (17) of \cite{ZhangDG},
is wrong.

Furthermore, comparing, in \cite{ZhangDG}, the first line of (17) with (2) for
$V=V_1V_2V_3$, we see that Zhang has set the two $V$'s equal
for all temperatures, implying the erroneous equality of
$H_y=\sum_{\tau=1}^{mn}\sigma^z_{\tau}\sigma^z_{\tau+m}$ and
Onsager's $A_m=\sum_{\tau=1}^{mn}\sigma^z_{\tau}
(\prod _{j=1}^{m-1}\sigma^x_{\tau+j}) \sigma^z_{\tau+m}$,
identifying $s=\sigma^z$ and $C=\sigma^x$ in (45) and (56) of \cite{Onsager}.
In fact, it is the typical error in most incorrect solutions of 3D Ising.
Since 1975, as a referee, I have rejected several manuscripts, in
which the 3D Ising model was incorrectly reduced
in a somewhat similar way to free fermions and I have commented
on one other such work \cite{ZDZ}; see also Section 6.2 of \cite{ZDZ2}.

There are more reasons to see that \cite{ZhangDG} is flawed.
The formula for the critical temperature (32),
\begin{equation}
\sinh(2\beta J)\sinh(2\beta J_1+2\beta J_2)=1,
\quad\beta=1/k_{\mathrm{B}}T,
\end{equation}
can lead to three different critical temperatures by just
rotating the lattice. Indeed, using ($J_1$, $J_2$, $J$) as a
permutation of (0.5, 1.0, 1.5), one obtains three critical
temperatures, with $\beta_c=1/k_{\mathrm{B}}T_c$ being 0.3503982204,
0.3046889317, or 0.2937911957, depending on $J$ being
0.5, 1.0, or 1.5. According to \cite{ZH}, however, these values are three upper bounds
on the true $\beta_c$, three lower bounds on the true $T_c$,
and (3) only provides the critical point asymptotically in the limit
$J\gg J_1,J_2\ge0$.

When $J_1=J_2=J=1$ one finds 0.3046889317, thus disagreeing with
about every result in the literature. The best value to date may be\linebreak
0.221654626(5) \cite{FXL}. Almost all calculations, using series,
Monte Carlo, finite-size extrapolations, renormalization group,
etc., point in the direction of 0.22. Only very few calculations
provide different numbers and one can quickly\linebreak locate obvious errors
in those calculations.

Furthermore, the critical exponents do not agree at all with the best
values in the literature, see, e.g.,\ \cite{FP,H,PV} and the references cited.
The critical exponent $\alpha=0$ found disagrees with $\alpha=0.11\cdots$
supported by the most reliable estimates in the literature.
The relation with the 2D Ising claimed would seem to imply $\beta=1/8$ and
$\eta=1/4$, which are far outside the values obtained using a series and Monte Carlo.

In conclusion, the errors in \cite{ZhangDG} cause the results to be wrong.

\eject
{\centering\LARGE Reply to  Symmetry 2023, 15, 375}
\\
\\
The published version of the comment \cite{PerkC} has appeared
back-to-back with a reply by Zhang Degang \cite{ZhangDG2}, which
contains more errors, besides the errors present in \cite{ZhangDG},
as I shall explain.

First of all, in his reply Zhang now agrees that in the thermodynamic
limit the results are independent of boundary conditions. This implies
that he now advertises (44), not (45), in \cite{ZhangDG} as the beginning
of the high-temperature series, also for periodic boundary conditions.

However, that (45) is the correct high-temperature series follows from the
rigorous treatment of a very general class of lattice models, for which the
proofs make use of the Banach space of interactions \cite{GMR,Ruelle2}.
In sections 3 and 4 of \cite{ZDZ2} a more explicit proof, restricted to 3D
Ising, has been outlined giving also a lower bound on the radius of convergence.

In his reply \cite{ZhangDG2} Zhang Degang claims that Eq.~(2) in \cite{PerkC} is wrong.
I agree with that statement, as it is intended to be wrong, being equivalent to the
error made by him. This is explicit in the reply \cite{ZhangDG2} on the second
to last line of page 1, where he writes
\begin{equation}
\sigma'^z_{p+am}=\sigma^x_{p+(a-1)m+1}\cdots\sigma^x_{p+am-1}\sigma^z_{p+am}.
\end{equation}
For the special value $a=1$ and multiplying by $\sigma^z_p$, we get
\begin{equation}
\sigma^z_p\sigma'^z_{p+m}=\sigma^z_p\sigma^x_{p+1}\cdots\sigma^x_{p+m-1}\sigma^z_{p+m}.
\end{equation}
Now $\sigma'^z_{p+m}$ and $\sigma^z_{p+m}$ must be equal, as they both measure
the same Ising spin being up or down ($+$ or $-$). Hence, (2) in \cite{PerkC} captures the
main error made by Zhang Degang, both in \cite{ZhangDG} and in \cite{ZhangDG2}.

Because of this error, Zhang Degang replaced $H_y$ in (2) of \cite{ZhangDG}
by $A_m$ in (16) and (17), in doing so wrongly converting the problem to one
solvable by free fermion methods. Indeed, using Kaufman's spinors $\Gamma_j$,
nowadays also called Majorana fermions, all $A_m$ and $G_m$ in this case
become quadratic in these  $\Gamma_j$'s, see (41) and (42) in \cite{PerkO}
and cited references\footnote{To compare with \cite{PerkO} requires a
rotation $\sigma_x\to\sigma_z$, $\sigma_z\to\sigma_x$, $\sigma_y\to-\sigma_y$.
This is often done in 2D Ising works since \cite{SML}, whose authors apparently
wanted notations to agree with their earlier paper
on the XY chain \cite{LSM}.}.

As Zhang Degang tried to ridicule Equation (2) in \cite{PerkC}, it may
be good to give some more detail how I got to it. To show that
$\sum_{p=1}^m \mathcal{A}_{p,1}=\sum_{p=1}^m A_{p,1}$ implies
$\mathcal{A}_{p,1}=A_{p,1}$, we can start with $m=2$. Let us write
the difference of the two-term sums in Kronecker product notation,
using $\sigma^z=(\begin{smallmatrix}1&0\\0&-1\end{smallmatrix})$,
$\sigma^x=(\begin{smallmatrix}0&1\\1&0\end{smallmatrix})$,
$\mathsf{1}_2=(\begin{smallmatrix}1&0\\0&1\end{smallmatrix})$,
and $\mathsf{X}=(\begin{smallmatrix}a&b\\c&d\end{smallmatrix})$
in place of $\sigma^x$ at first. Then the difference of the two
two-term sums for $m=2$ becomes
\begin{eqnarray}
&&(\sigma^z\otimes\mathsf{X}\otimes\sigma^z\otimes\mathsf{1}_2+
\mathsf{1}_2\otimes\sigma^z\otimes\mathsf{X}\otimes\sigma^z)\nonumber\\
&&\qquad-(\sigma^z\otimes\mathsf{1}_2\otimes\sigma^z\otimes\mathsf{1}_2+
\mathsf{1}_2\otimes\sigma^z\otimes \mathsf{1}_2\otimes\sigma^z)= \nonumber\\
&&\hspace{-0.2in}\setcounter{MaxMatrixCols}{16}
\left[\begin{smallmatrix}\\
2a-2&0&b&0&b&0&0&0&0&0&0&0&0&0&0&0\\
0&0&0&-b&0&b&0&0&0&0&0&0&0&0&0&0\\
c&0&\!\!-a+d&0&0&0&-b&0&0&0&0&0&0&0&0&0\\
0&-c&0&\!\!\!\!-a+2-d&0&0&0&-b&0&0&0&0&0&0&0&0\\
c&0&0&0&\!\!\!\!\!\!-a+d&0&-b&0&0&0&0&0&0&0&0&0\\
0&c&0&0&0&\!\!d-2+a&0&b&0&0&0&0&0&0&0&0\\
0&0&-c&0&-c&0&\!\!\!\!\!-2d+2&0&0&0&0&0&0&0&0&0\\
0&0&0&-c&0&c&0&0&0&0&0&0&0&0&0&0\\
0&0&0&0&0&0&0&0&0&0&b&0&-b&0&0&0\\
0&0&0&0&0&0&0&0&0&-2a+2&0&-b&0&-b&0&0\\
0&0&0&0&0&0&0&0&c&0&\!\!\!\!d-2+a&0&0&0&b&0\\
0&0&0&0&0&0&0&0&0&-c&0&\!\!\!\!a-d&0&0&0&b\\
0&0&0&0&0&0&0&0&-c&0&0&0&\!\!-a+2-d&0&-b&0\\
0&0&0&0&0&0&0&0&0&-c&0&0&0&\!\!\!\!a-d&0&b\\
0&0&0&0&0&0&0&0&0&0&c&0&-c&0&0&0\\
0&0&0&0&0&0&0&0&0&0&0&c&0&c&0&\!\!2d-2
\end{smallmatrix}\right]\nonumber\\
&&\qquad=0.\end{eqnarray}
The only solution is $a=d=1$, $b=c=0$, or
$\mathsf{X}=\mathsf{1}_2$. It is false for
$\mathsf{X}=\sigma^x$ with nonzero off-diagonal elements
$b=c=1$ and similarly false when $m>2$, implying that
$H_y\ne A_m$ for all $m$. This shows again that (16)
and (17) in \cite{ZhangDG} are inconsistent with (2) therein.

We may also write the difference of the two-term sums as
\begin{equation}
\mathsf{M}\otimes\mathsf{1}_2+\mathsf{1}_2\otimes\mathsf{M}=0,
\quad\mathsf{M}\equiv\sigma^z\otimes(\sigma^x-\mathsf{1}_2)\otimes\sigma^z.
\end{equation}
Even dropping the definition of the $8\times8$ matrix $\mathsf{M}$
it is easily shown that the equation solves as $\mathsf{M}=0$.
This line of thought can be generalized to $m>2$. The case $m=2$
generalizes most easily to the case $m=2^n$, taking $M$ to be the
sum of the first $2^{n-1}$ terms and showing that this $M$
vanishes and then repeating for the first $2^{n-2}$ terms, etc.

Finally, it may be noted that the use of screw boundary conditions
in the three-dimensional Ising model had already been
advertised in section 6 of \cite{ZDZ2}. Also, in (26) and (27)
of \cite{ZDZ2}, the three factors of the transfer matrix
are correctly expressed in terms of Kaufman spinors.
Here (26) is clearly not a fermionic Gaussian,
see also the comparison of (3) and (4) in \cite{ZDZ}.

\eject

\end{document}